 \documentclass[final,5p,times,twocolumn]{elsarticle}

\usepackage{amssymb}
\usepackage{amsthm}
\usepackage{amsmath}
\usepackage{tabularx}
\usepackage{float}
\usepackage[hyphens]{url}
\usepackage{subfigure}

\usepackage{lineno}

\journal{Nuclear Physics B}

\begin{document}

\begin{frontmatter}



\title{{Interferometer-based high-accuracy white light measurement of neutral rubidium density and gradient at AWAKE}}

 \author[label1,label2,label3]{F. Batsch}
 \author[label2]{M. Martyanov}
 \author[label2]{E. Oez} 
 \author[label2]{J. Moody}
 \author[label1]{ E. Gschwendtner}
 \author[label2]{ A. Caldwell}
 \author[label1,label2]{ P. Muggli}
 \address[label1]{CERN, Geneva, Switzerland}
 \address[label2]{Max Planck Institute for Physics, Munich, Germany}
 \address[label3]{Technical University Munich, Munich, Germany}

\begin{abstract}
The AWAKE experiment requires an automated online rubidium (Rb) plasma density and gradient diagnostic for densities between 1 and $10\cdot10^{14}$ cm$^{-3}$. A linear density gradient along the plasma source at the percent level may be useful to improve the electron acceleration process. Because of full laser ionization of Rb vapor to Rb$^{+}$ within a radius of 1 mm, the plasma density equals the vapor density. We measure the Rb vapor densities at both ends of the source, with high precision using, white light interferometry. At either source end, broadband laser light passes a remotely controlled Mach-Zehnder interferometer built out of single mode fibers. The resulting interference signal, influenced by dispersion in the vicinity of the Rb D1 and D2 transitions, is dispersed in wavelength by a spectrograph.
 Fully automated Fourier-based signal conditioning and a fit algorithm  yield the density with an uncertainty between the measurements at both ends of 0.11 to 0.46 $\%$ over the entire density range. These densities used to operate the plasma source are displayed live in the control room. 
\end{abstract}

\end{frontmatter}


\section{Introduction}
\label{Intro}
\begin{figure}[h!]
\centering
\includegraphics[height=5.4cm]{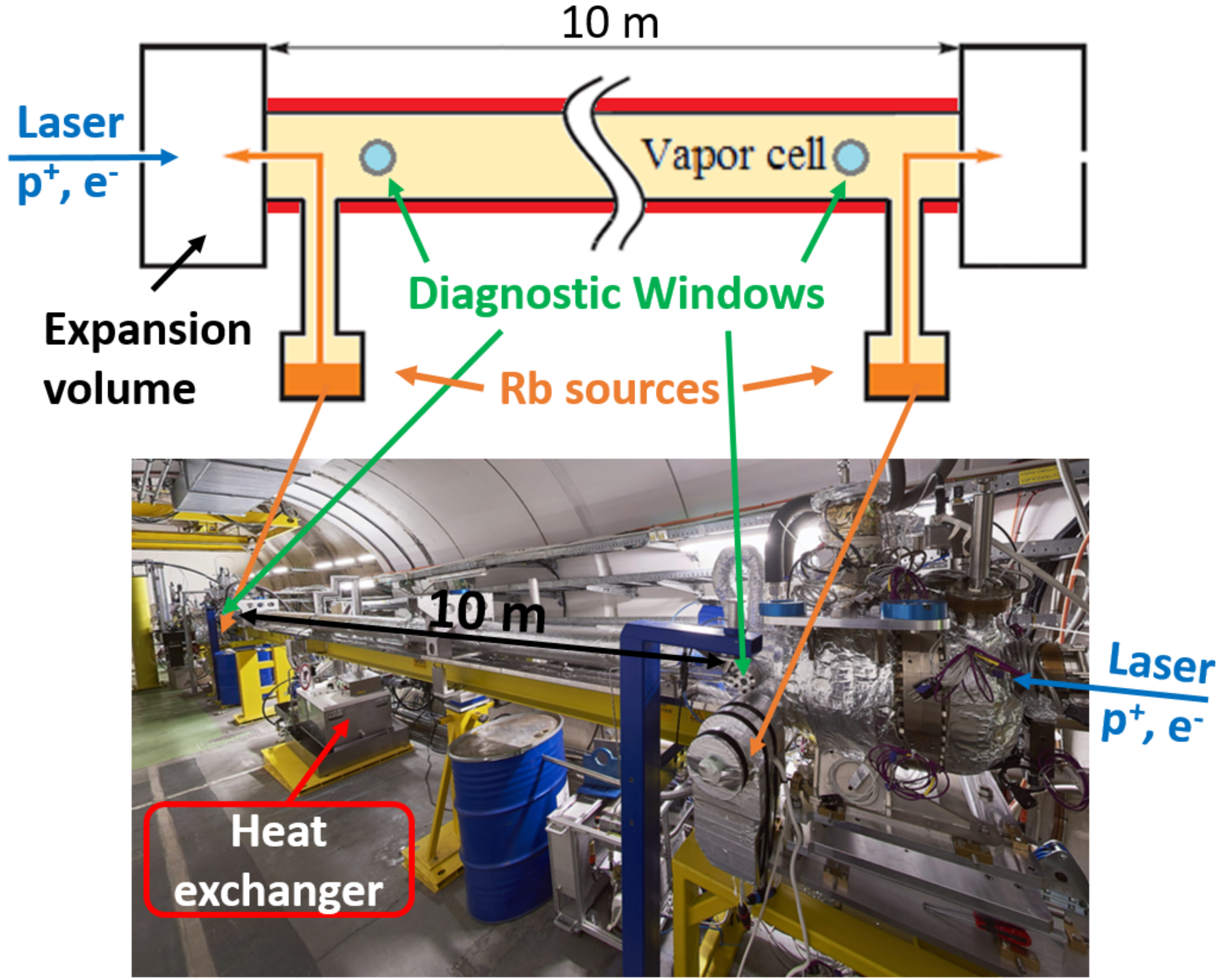}
\caption{Top: Schematic of the Rb vapor source showing the 10 m long pipe surrounded by the heat exchanger (red), two Rb reservoirs (orange) providing the Rb vapor and 2 diagnostic viewports near the source ends. Bottom: Photo of the AWAKE vapor source. The blue posts at each end are supports for the interferometer optics.}
\label{fig:rbcelldrawing}
\end{figure}

The AWAKE project at CERN is a proof-of-concept experiment that uses a proton bunch for particle beam driven plasma wakefield acceleration of electrons \cite{awakecollaboration,GSCHWENDTNER201676,PathtoAWAKE}. The goal is to reach energies on the scale of several GeV using coherently driven plasma waves with acceleration gradients $>$ 1 GeV/m \cite{readiness}. The entire process, i.e. modulating the 12 cm long ($\sigma_z$), 400 GeV proton bunch \cite{PathtoAWAKE} by seeded self-modulation (SSM) \cite{readiness,smi} into micro bunches, wakefield creation and electron acceleration, happens in a 10 m long, 4 cm diameter rubidium (Rb) vapor source \cite{IPAC,Statusreport2016,gennadyplasmacell}, depicted in Fig. \ref{fig:rbcelldrawing}. 
 At each end, a flask with separately controlled electrical heaters is filled with Rb, providing Rb vapor densities up to $1\cdot10^{15}$ cm$^{-3}$. The baseline density is $n_{Rb}=7\cdot10^{14}$ cm$^{-3}$ \cite{Statusreport2016}. A fluid heat exchanger with temperature-stabilization surrounds the source and ensures a high temperature and vapor density uniformity ($<$ 0.2 $\%$, \cite{IPAC}). An intense laser pulse ionizes the Rb vapor (first e$^-$ of each Rb atom), forming a 2 mm diameter plasma along the source with equal density and uniformity. By setting different temperatures in the downstream and upstream flasks, a linear vapor / plasma density gradient along the source can be set. Beside the density uniformity, the absolute vapor density and a possible gradient along the source influence the acceleration process \cite{Petrenko}. The absolute density determines the proton bunch modulation frequency. Density gradients on the order of +1 to +10 $\%$ (i.e. the density increases along the 10 m pipe in direction of the beam) can affect the e$^-$ acceleration in a positive way \cite{Petrenko}.\\
  We determine the plasma density and gradient by measuring the Rb vapor density through diagnostic windows located at each of the source ends (see Fig. \ref{fig:rbcelldrawing}) using a Mach-Zehnder interferometer and white light interferometry \cite{THill,MasterThesis}. To ensure a sufficiently high accuracy in gradient determination, we aim for an uncertainty in measuring the densities at both source ends to better than 1 $\%$. To operate the vapor source remotely, from the control room, while ensuring the required densities and gradients, the diagnostic must allow for a fully automated and remote-controlled operation and provide online density values. The analysis to determine the densities is based on Fourier signal conditioning and on a fitting algorithm analyzing zero-crossings. The diagnostic and the signal analysis are described hereafter.

\section{The diagnostic}
\label{Physics} 
\begin{figure}[h!]
\centering
\includegraphics[height=3.9cm]{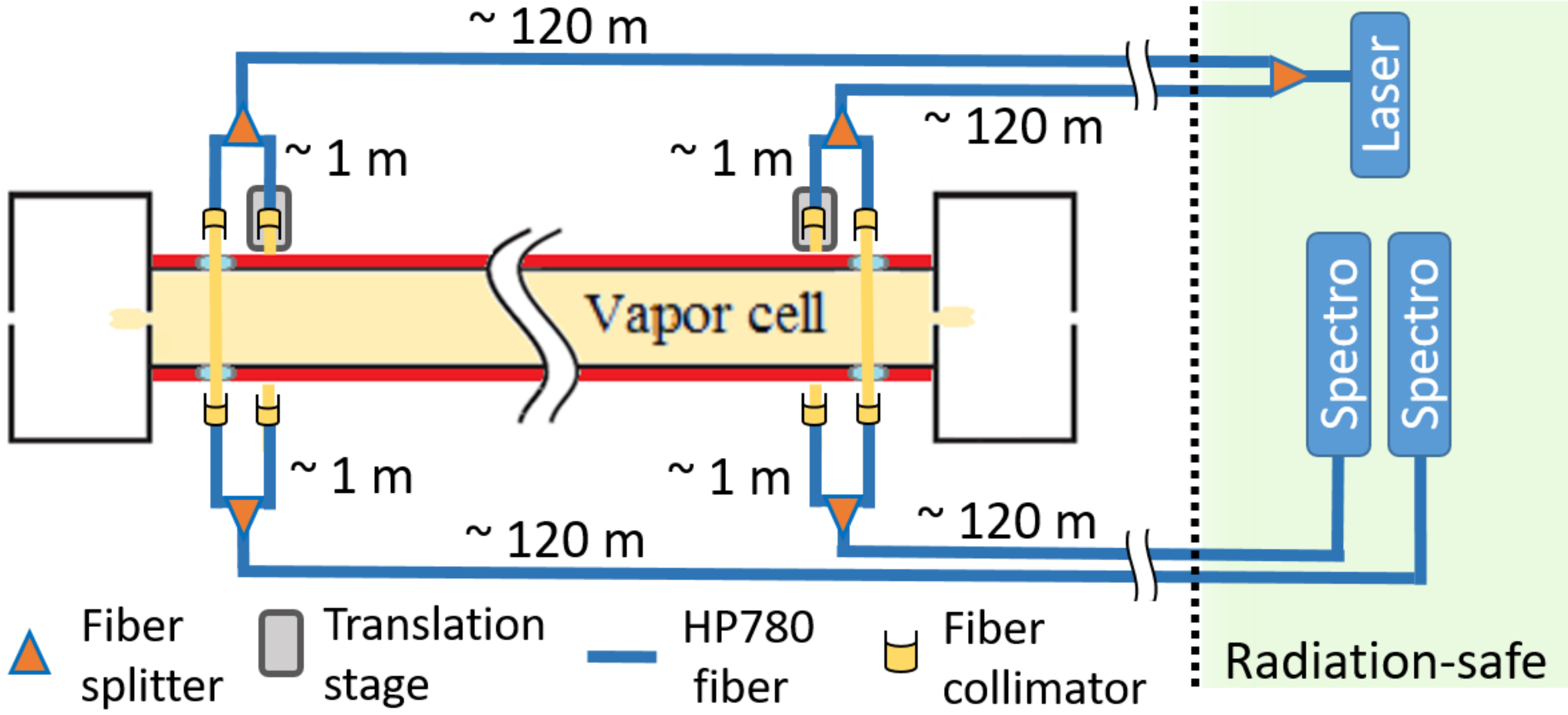}
\caption{Top view schematic of the fiber-based Mach-Zehnder interferometer assembly. From the light source, located in a radiation-safe area, 120 m fibers transport the light to the source ends, where the interferometers are formed by fiber splitters. The light traverses the source through the diagnostic windows, each reference arm is equipped with a translation stage to adjust its length. Equal length fibers transport the interfered light signal back to two fiber spectrographs.}
\label{fig:interfdrawing}
\end{figure}

\begin{figure}[h!]
\centering
\includegraphics[height=6cm]{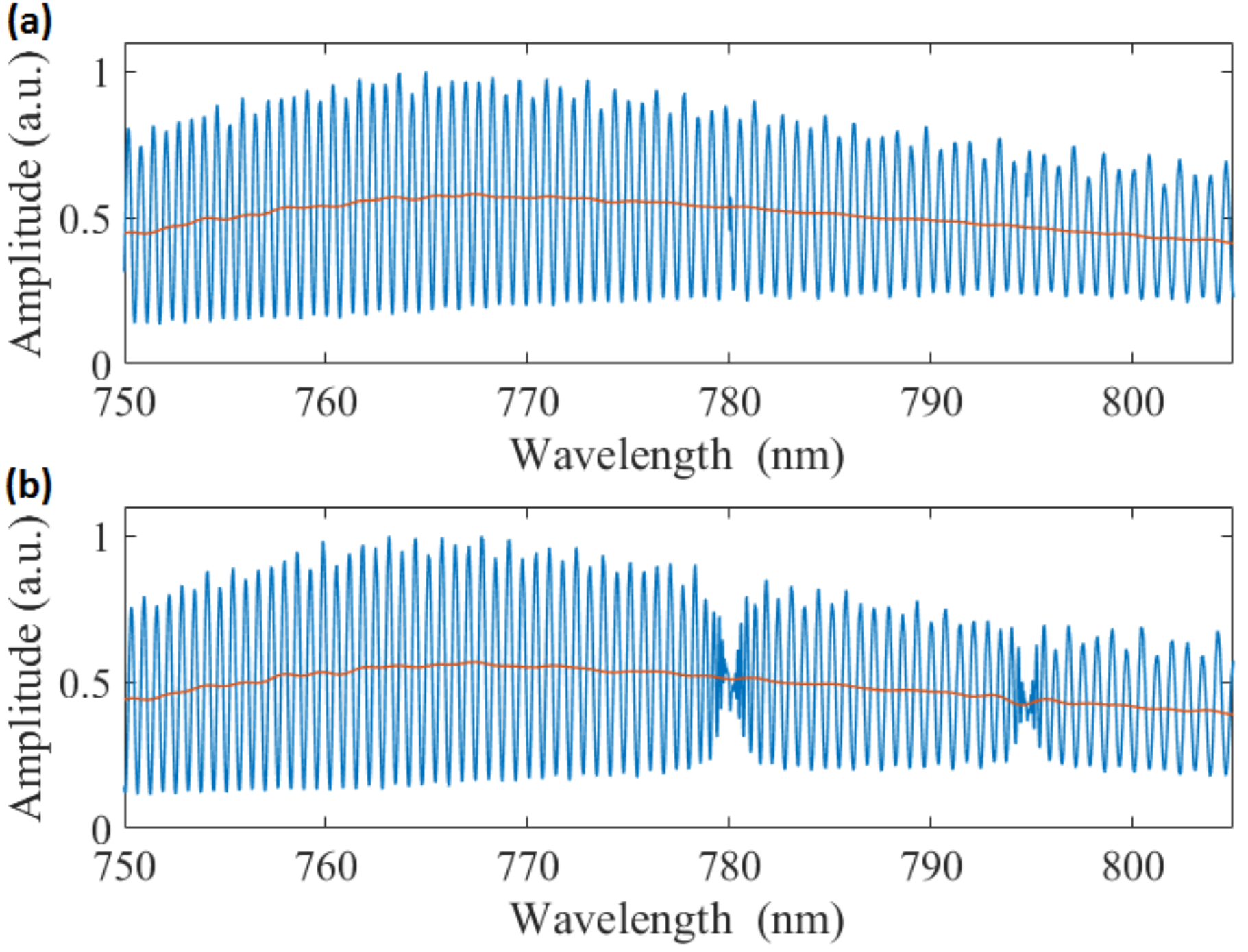}
\caption{(a) Interference pattern without Rb in the Rb vapor source. The oscillation shows nearly constant period.   \ (b) Interference pattern with $n_{Rb}=1.365\cdot10^{14}$ cm$^{-3}$  in the source. Around the transition lines (D1 at 795 nm and D2 at 780 nm), the period changes in a density-dependent way. The red lines represent the offset of the oscillation (see later).}
\label{fig:intercurves}
\end{figure}

The technique exploits the fact that alkali metals, such as Rb, have atomic transitions from the ground state to the first exited state in the optical wavelength range. Rubidium has two such lines, at 780.03 nm (D2) and 794.76 nm (D1) \cite{rbtransitionnumber,steck87}. In the vicinity of these transitions, its optical properties change with wavelength (dispersion) and Rb density. This density-dependent change in the index of refraction for each wavelength results in an interference pattern that changes with density. We measure it by sending coherent white light in a fiber-based Mach-Zehnder interferometer and through the Rb vapor. \\
This setup, depicted in Fig. \ref{fig:interfdrawing}, includes a white light laser (NKT SuperK COMPACT, 240 - 2000 nm spectrum) as light source, located in a radiation-safe area.
Wavelengths between 700-900 nm are then coupled into two single-mode optical fibers. These $\approx$ 120 m long fibers lead to each vapor source end. A fiber splitter forms the two arms of the Mach-Zehnder interferometer. One is called Rb arm in the following and guides the light to the diagnostic window. At the fiber end, a fiber collimator forms a parallel light beam that passes the Rb vapor transversely. A second fiber collimator re-couples the light into the fiber. The second arm, called reference arm, is a replica of the Rb arm and is located below the vapor source. Its free-space section (length equal to that of the Rb arm) contains a translation stage to adjust the path length difference between the two arms. Another fiber splitter recombines the light from both arms. The interfered signals from both interferometers propagate over a second pair of $\approx$ 120 m fibers (equal length), back to the  radiation-safe area. There, two Ocean Optics HR4000 fiber spectrographs disperse the signals in wavelength with a resolution of 0.063 nm \cite{MasterThesis}. Figure \ref{fig:intercurves} shows the resulting interference patterns for the cases of no Rb and Rb vapor with a density of $1.365\cdot10^{14}$ cm$^{-3}$ in the source.

\section{Density calculation}
\label{fitcode}
\begin{figure}[h!]
\centering
\includegraphics[height=5.2cm]{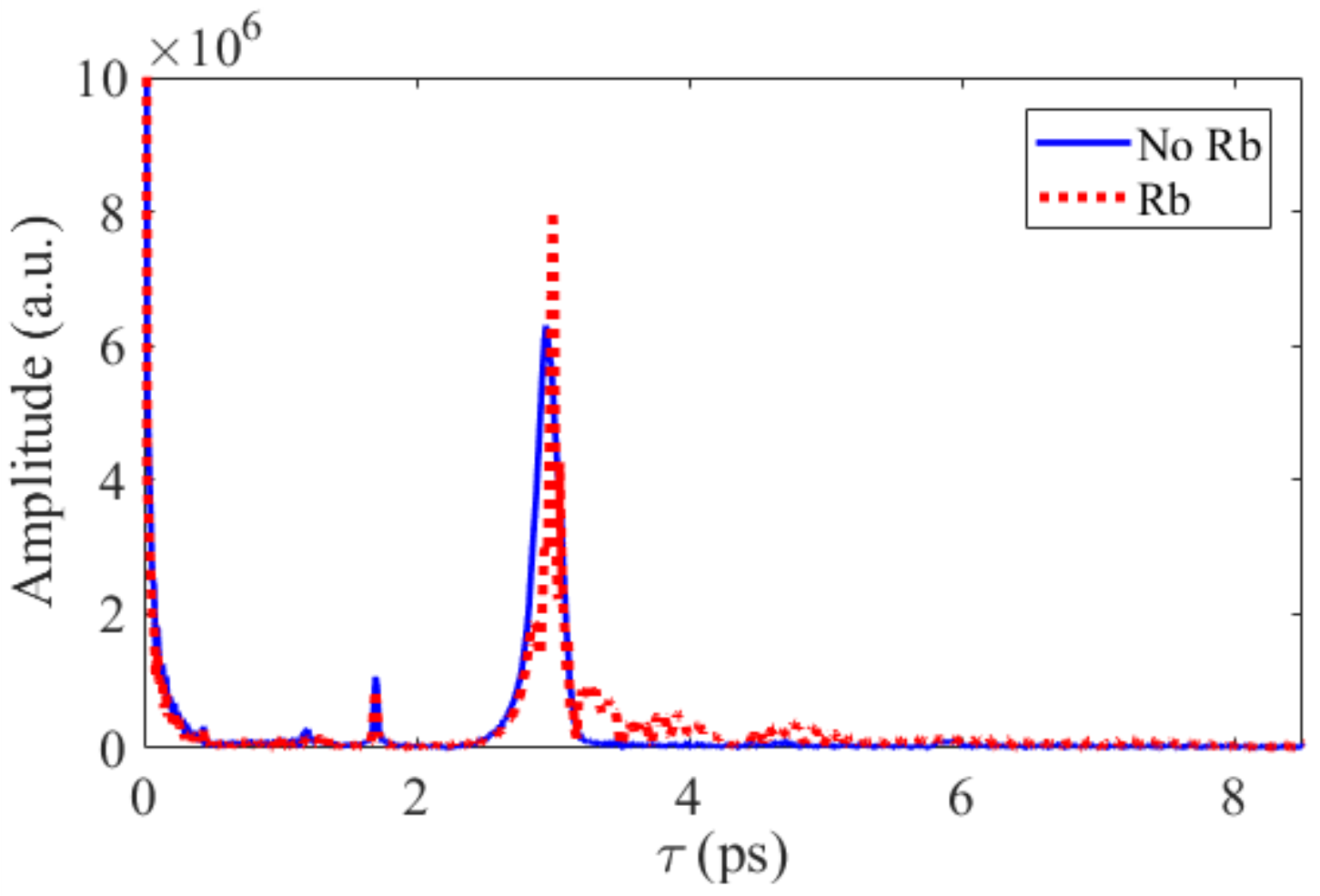}
\caption{Power spectra (i.e. absolute value of the Fourier transform) of interference spectra in case of no Rb vapor in the source (blue line) and in case of Rb vapor present ($n_{Rb}=1.365\cdot10^{14}$ cm$^{-3}$) (red).}
\label{fig:fft}
\end{figure}

\begin{figure}[h!]
\centering
\includegraphics[height=5cm]{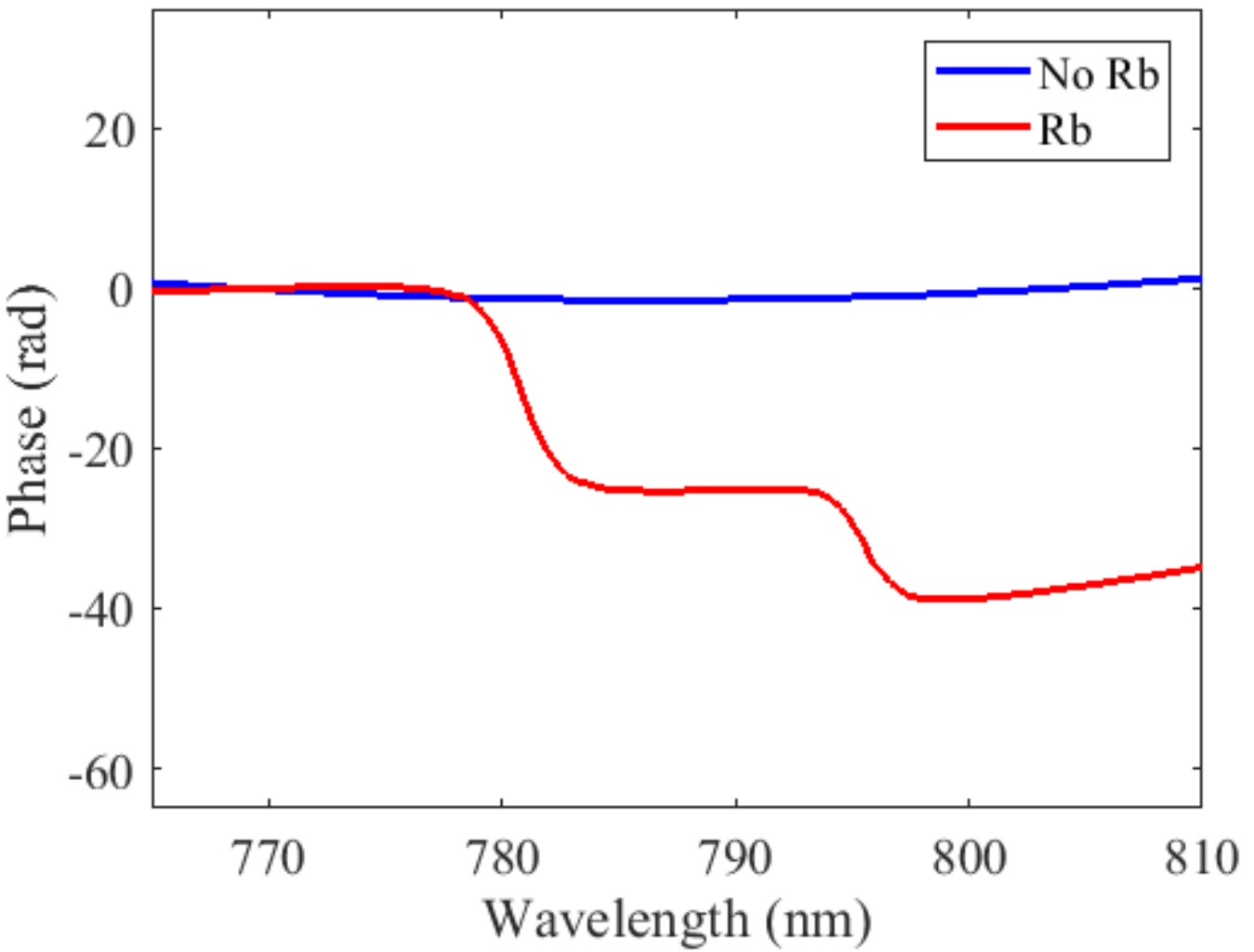}
\caption{Measured phase difference in case of no Rb vapor (blue line) and in case of Rb vapor present ($n_{Rb}=2.092\cdot10^{14}$ cm$^{-3}$) (red). The difference between both lines corresponds to $\Delta\Phi$.}
\label{fig:phases}
\end{figure}

\begin{figure}[h!]
\centering
\subfigure
{
\includegraphics[height=4.6cm]{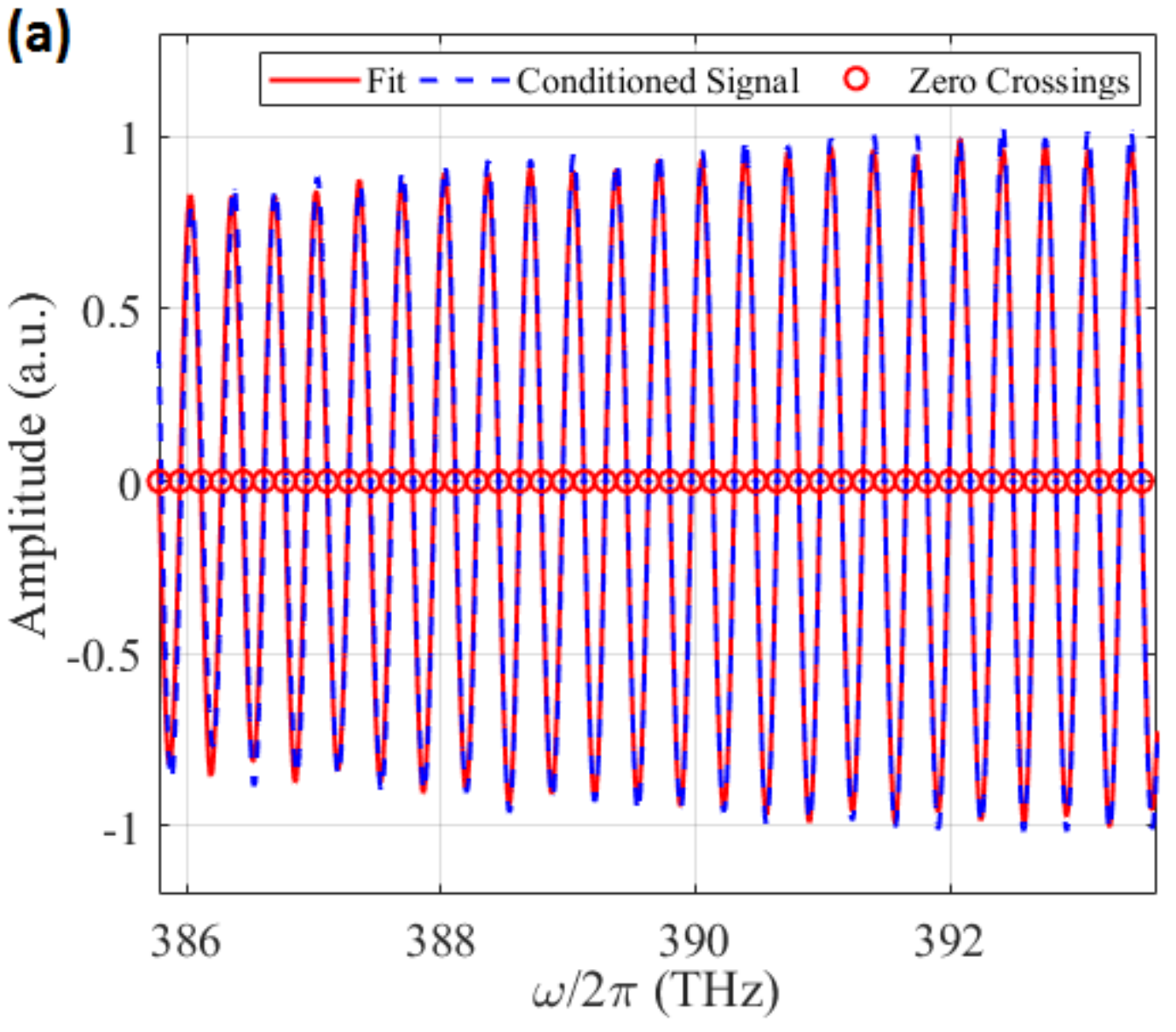}
\label{fig:fit}}
\hskip2mm
\subfigure
{
\includegraphics[height=4.6cm]{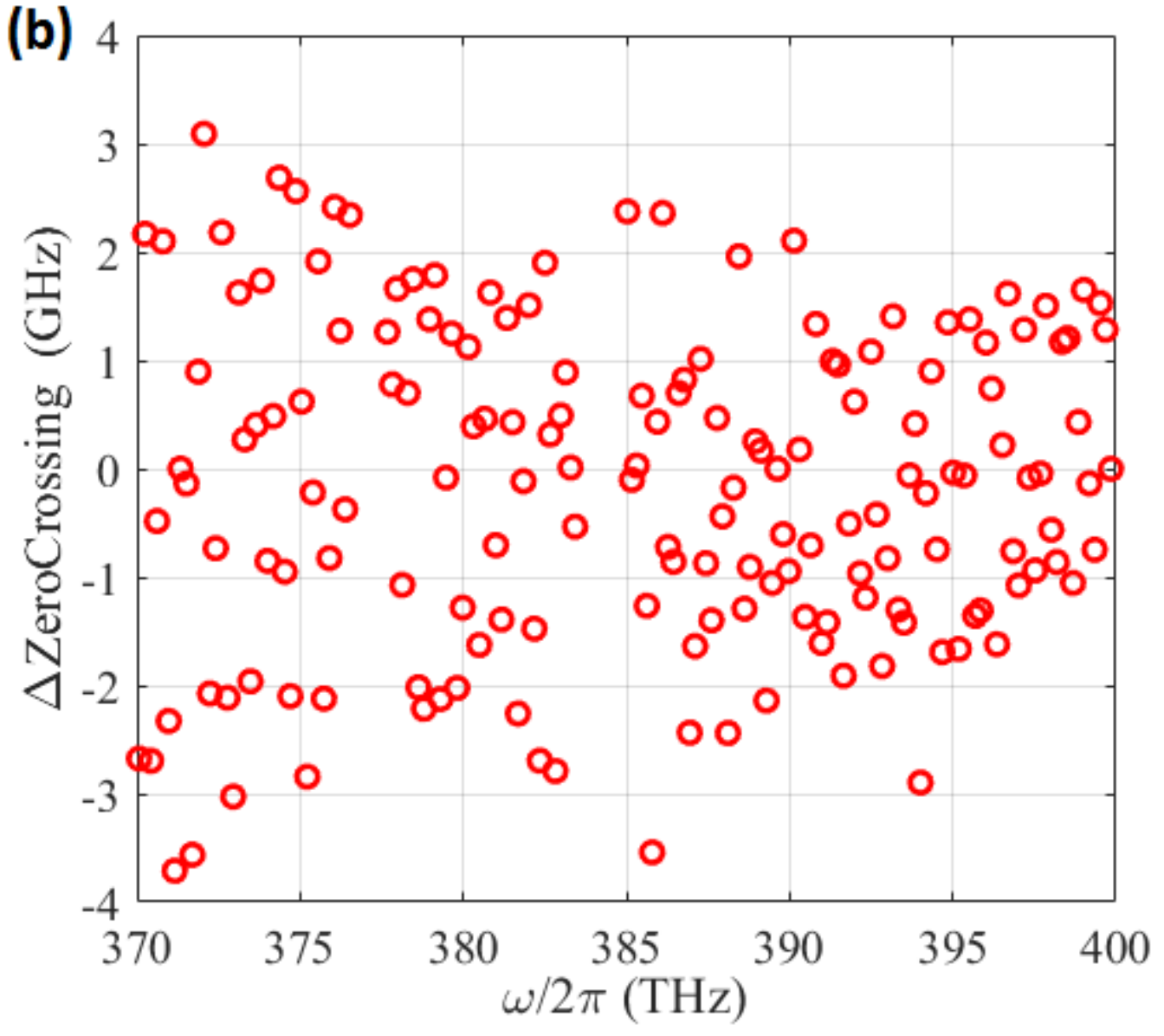}
\label{fig:crossings}
 }
 \caption{(a) Plot of the conditioned signal (blue), the zero crossings (red circles) and the fit(red line) vs. frequency ($\omega/2\pi$) for the same spectrum as shown in Fig. \ref{fig:intercurves} (b) in the vicinity of the D2 transition. \ (b) Plot of the difference between the zero-crossings of the signal and the fit  vs. frequency (range now covering both transitions).}
\end{figure}

In interferograms such as those of Fig. \ref{fig:intercurves}, the interference pattern is given by
\begin{equation}
I_{tot}(\lambda)=I_1(\lambda)+I_2(\lambda)+2\sqrt{I_1(\lambda)I_2(\lambda)}\cdot \cos(\Delta\Phi(\lambda)) \ ,
\label{eq:fitequation1}
\end{equation}
 where $I_{1,2}(\lambda)$ are the light intensities in each interferometer arm at wavelength $\lambda$, $I_{1}(\lambda)+I_{2}(\lambda)$ is the oscillation offset (see red line in Fig. \ref{fig:intercurves}) and $\Delta\Phi(\lambda)$ the phase difference between the arms. Changing from wavelength $\lambda$ to frequency $\omega$, this phase difference is described by
 \begin{align}
 \begin{split}
 \Delta \Phi(\omega)=& \ k\eta(\omega)\cdot(l_{F1}-l_{F2})+k\cdot(l_1-l_2)+(\Phi_{01}(\omega)-\Phi_{02}(\omega)) \\
 & +kL(\eta_{Rb}(\omega)-1) \ .
 \end{split}
 \end{align}
 Here, $k=\omega/c$ is the wavenumber in vacuum, $\eta(\omega)$ the fiber's index of refraction, $l_{F1,2}$ the fiber lengths of each arm, $l_{1,2}$ the path lengths in free space outside the source or fibers, $\Phi_{01,2}$ the phase of the light in each arm, $L$ the length of the Rb vapor column through which the light propagates and $\eta_{Rb}$ the index of refraction of the Rb vapor. Taylor-expanding the fiber's index of refraction around center frequency $\omega_0$, one can rewrite the phase difference as 
 \begin{align}
 \begin{split}
\Delta\Phi(\omega)&=\left[\frac{1}{2}\alpha\Delta\omega^2+\beta\Delta\omega+\Delta l+\Delta\Phi_0\right]+\left[\frac{\omega}{c}L(\eta_{Rb}(\omega)-1)\right] \\ &:= [A]+[B] \ .
\end{split}
\label{eq:phase}
\end{align}
 Here, the  path length difference $\Delta l=l_1-l_2$, $\Delta\Phi_0=\Delta\Phi_{01}-\Delta\Phi_{02}$ and $\alpha$, $\beta$ include all frequency-independent terms. The first bracket contains the phase terms that are density-independent and we call it [A]. The second term ([B]) contains the terms that depend on the Rb vapor density $n_{Rb}$ through $\eta_{Rb}=\sqrt{1+\chi_e}$ and
\begin{equation}
\chi_e=\frac{e^2n_{Rb}}{\epsilon_0 m_e}\sum_{i=1,2}\frac{f_{i}}{(\omega_{i}^2-\omega^2)^2-i\gamma_{i}^2\omega^2} \ .
\label{eq:chi}
\end{equation}
Here, $\chi_e$ is the electric susceptibility, $i$ the index of the transitions (D1 and D2), $\omega_{i}=(2\pi c)/ \lambda_i$ the transition frequencies, $e$ the electron charge, $ \epsilon_0$ the vacuum permittivity, $m_e$ the electron mass, $f_{i}$ the transitions oscillator strength and $\gamma_{i}$ its natural lifetime \cite{MasterThesis,steck87,NIST}. Doppler broadening is taken into account by correcting $\chi_e$ accordingly using a Rb temperature of $200^{\circ}$C for all densities. However,  we exclude a frequency range of width 0.15 THz around each transition line from the analysis because this range includes, in addition to the absorption lines, the not resolvable short-period oscillations (see Fig. \ref{fig:intercurves}(b)).  \\
As described by these formulas, the effect of the Rb vapor on the phase difference is proportional to the density-length product $n_{Rb}L$ (which appears after a binomial expansion of $\sqrt{1+\chi_e}$ to first order). However, we treat $L$ as constant. The heat expansion factor of steel is negligible ($\sim10^{-5}m/m \ K$) and equal for both vapor column lengths, meaning that it does not affect the gradient measurement. To ensure low systematic error in density measurement, $L$ was measured with 0.02 $\%$ accuracy using a micrometer. \\
Before calculating the density from the phase shift induced by the Rb vapor ([B] in Eq. \ref{eq:phase}) using a fitting algorithm, the spectra must be normalized and the offset ($I_{1}(\lambda)+I_{2}(\lambda)$) must be removed for the fitting algorithm (since the vapor density length product is contained only in the argument of the cosine in Eq. \ref{eq:fitequation1}). In addition, we extract the signal's envelope function which is required for the fit. For these steps, we use Fourier-based signal conditioning. Figure \ref{fig:fft} shows the absolute value of the fast Fourier transform (FFT) of the spectra shown in Fig. \ref{fig:intercurves}.
 The small oscillation times $\tau \ (<2$ ps) represent the oscillation offset and high-frequency noise. Setting this part of the Fourier spectrum to zero and applying an inverse Fourier transform of the remaining spectrum removes the offset (and noise), i.e. centers the oscillation around the horizontal  axis. Further, the FFT spectrum shows a prominent peak (here at $\tau_{peak}\approx3$ ps). In case of no Rb vapor in the source (blue line), it represents the oscillation with constant period, determined by $\Delta l$ (for Fig. \ref{fig:intercurves}, $\Delta l\approx9$ mm). Shifting this peak to zero and taking the absolute value of its inverse Fourier transform gives the oscillation's envelope function. Determining the phase of this inverse Fourier transform with respect to $\omega_0$ (set to 390 THz $\hat{=}$ 770 nm) gives the phase difference $\Delta\Phi(\omega)$. Note that large $\tau$ values (i.e. $\tau > \tau_{peak} + 10$ ps) are zeroed before these steps as well, in order to remove the non-physical $\tau$ values which could possibly lead to an incorrect centering of the oscillation. We call this entire process signal conditioning. With Rb, the curve (red dashed line) looks similar, except around the prominent peak. The changing period around the Rb transition wavelength in the interferogram leads to a broader peak and a different $\Delta\Phi(\omega)$. \\
 After the conditioning, one determines the density using a spectrum where $n_{Rb}=0$ (i.e. [B]=0) to obtain [A] and calculates $\Delta\Phi(\omega)$ for both cases (called $\Delta\Phi_{NoRb}$ and $\Delta\Phi_{Rb}$; note that all other parameter such as $\Delta l$ must be equal). Measured examples for these phase terms are depicted in Fig. \ref{fig:phases}.\\
 The comparison of $\Delta\Phi_{Rb}-\Delta\Phi_{NoRb}$ with the expression for [B] gives a first estimate for the density that is used as a starting value for the final fit. This final fit in the next step minimizes the distance between the zero-crossings of the conditioned signal and the curve given by the formula one obtains by multiplying $cos(\Delta\Phi)$ with the envelope function calculated in the conditioning process. The cosine term is obtained by substituting Eq. \ref{eq:chi} and the Rb density start value in Eq. \ref{eq:phase}. The terms $\alpha$, $\beta$ and $(\Delta l+\Delta\Phi_0)$ are kept as fitting parameters since the lengths, i.e. also the initial phases change slightly due to vibrations. To obtain their starting values, one fits $\Delta\Phi_{NoRb}$ with a second-order polynomial. Figure \ref{fig:fit} shows the conditioned signal (in blue), the zero-positions (red circles) and the fit (red line) for the case of Rb vapor in the source with $n_{Rb}=1.365\cdot10^{14}$ cm$^{-3}$. The plotted frequency range covers only one side of the D2 line ($\omega_2/2\pi=384$ THz) for a better visibility of the oscillations. The fit matches the data in shape (envelope) and zero-crossing position. The difference between the zero-crossings of the conditioned signal and the fit is a measure for the goodness of the fit. It is plotted in Fig. \ref{fig:crossings} for a wider frequency range covering both transition frequencies ($\omega_1/2\pi=377$ THz).  The differences are within $\pm$ 4 GHz, which is below the spectrograph resolution, meaning that the fit matches the data. \\
During the experiment, this algorithm  calculates in a fully automated way (every 10 seconds) the densities (duration of the calculation: $\approx1$ sec) for each source end. Using these density values, one calculates the gradient over 10 m $\Delta n_{Rb}=(n_{Rb,2}-n_{Rb,1})/n_{Rb,1}$, where $n_{Rb,1}$ is the upstream value and $n_{Rb,2}$ the downstream value. These density and gradient values are then displayed live in the control room.

\section{Diagnostic Operation and Accuracy}
\label{opresults}
\begin{figure}[h!]
\centering
\includegraphics[height=6.2cm]{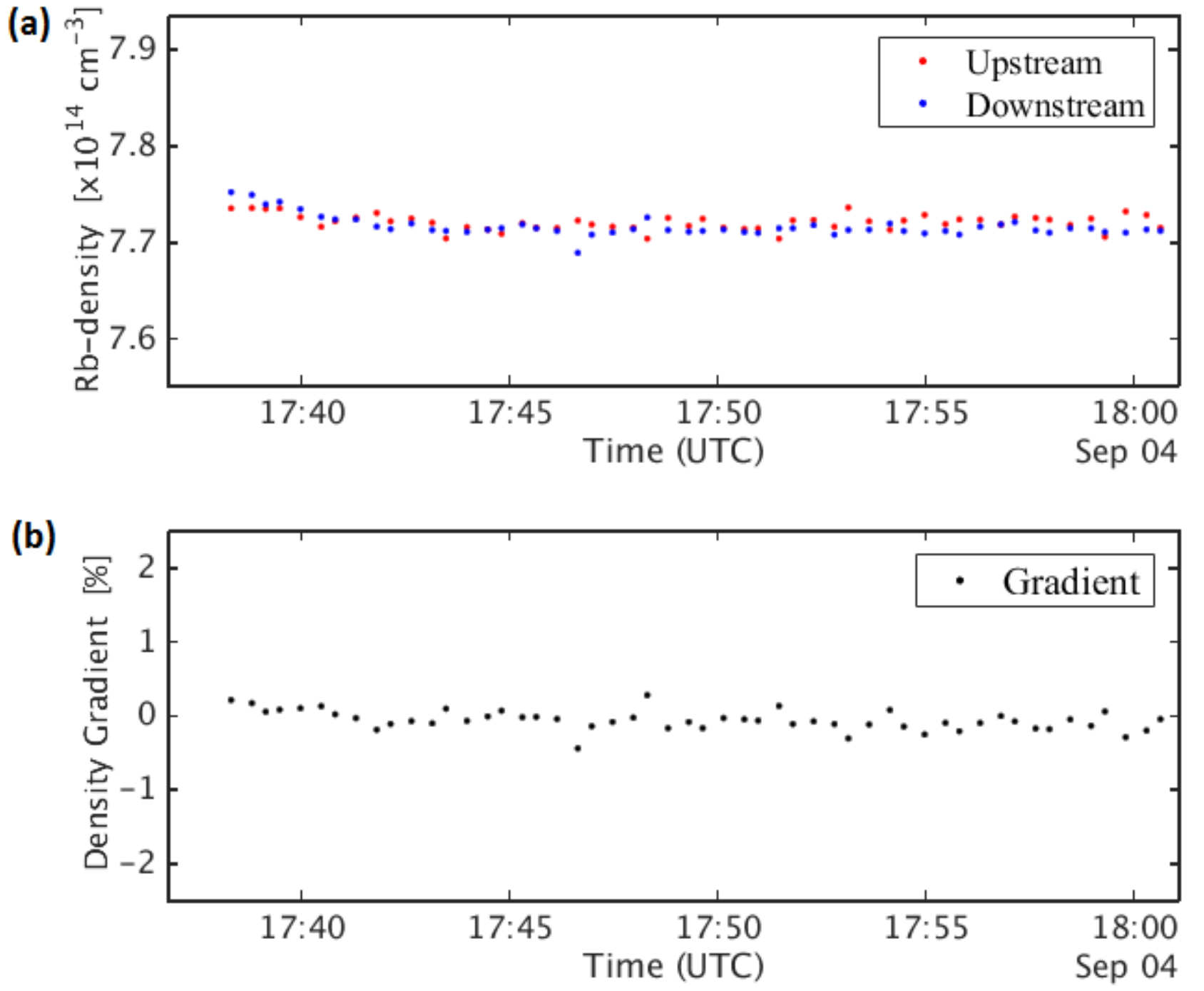}
\caption{Plot of (a) Rb vapor density vs. time (UTC) and (b) resulting gradient vs. time (UTC) for both source ends (upstream in red, downstream in blue).}
\label{fig:constdens}
\end{figure}

\begin{figure}[h!]
\centering
\includegraphics[height=6.2cm]{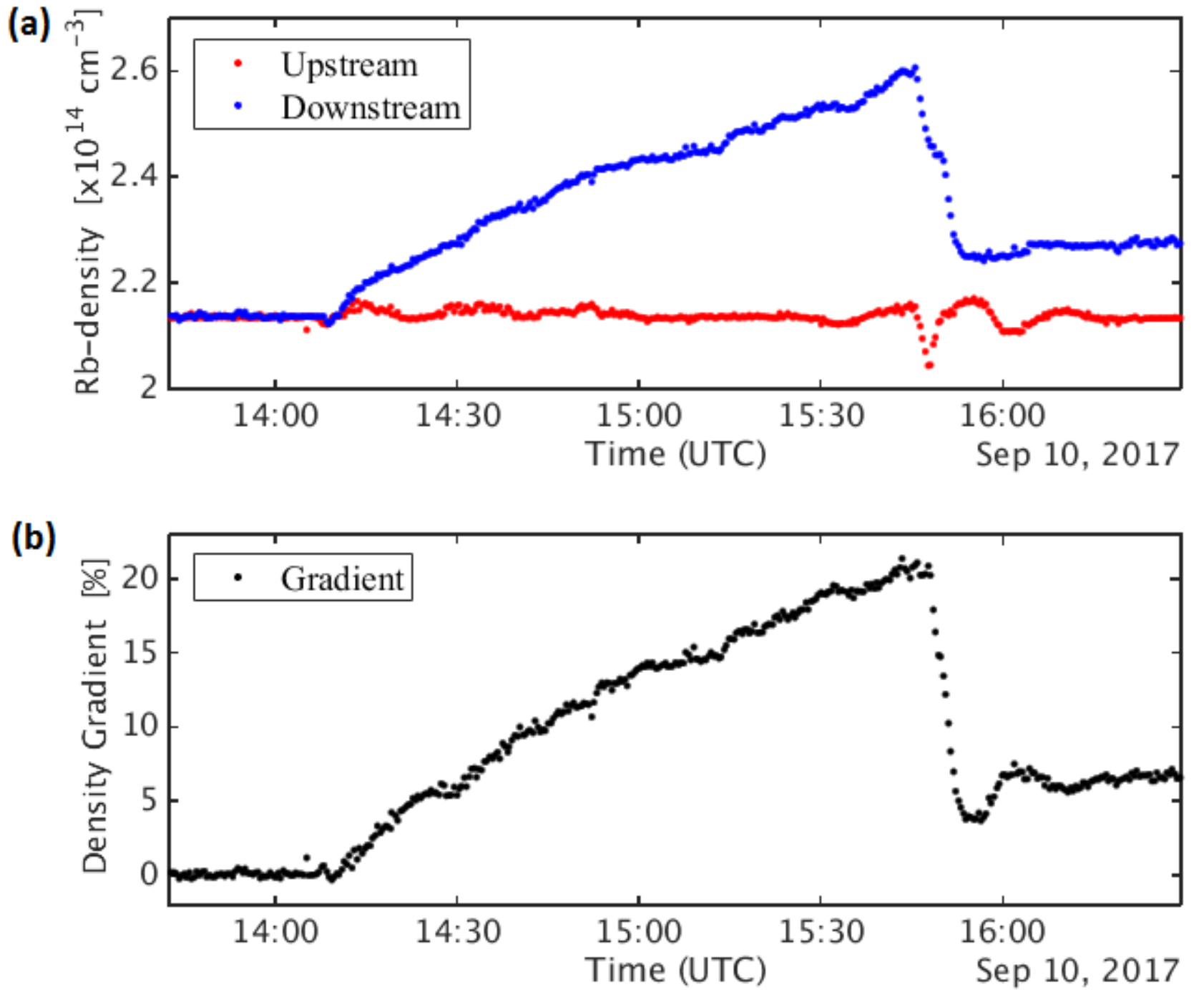}
\caption{(a) Plot of Rb vapor density vs. time (UTC) during a gradient scan. The downstream density (blue) is increases and then lowered again while the density upstream (red) is kept constant. \ (b) The resulting density gradient over 10 m along the vapor source vs. time (UTC). Positive gradients indicate that the density is higher downstream.}
\label{fig:gradscan}
\end{figure}

The diagnostic has three main tasks for the Rb vapor source operation. First, it is used to characterize the correlation between the temperature set in the Rb reservoirs and the Rb vapor density in the source. Due to the fact that it is an open system (see Fig. \ref{fig:rbcelldrawing}), a calculation of the density from temperature / a vapor pressure curve and the determination of the systematic uncertainty of one density measurement are not possible \cite{gennadyplasmacell}. Temperatures between $146.0^{\circ}$C and $215.0^{\circ}$C in the Rb flasks lead to densities between $n_{Rb}=3.09\cdot10^{13}$ cm$^{-3}$ and $n_{Rb}=10.70\cdot10^{14}$ cm$^{-3}$. The systematic uncertainty of one diagnostic was checked previously \cite{MasterThesis}, where the same setup and a comparable analysis algorithm was used. A closed metallic cube with two viewports that was immersed in an oil bath with temperature stabilization (0.1 $^{\circ}$C uncertainty) and calibrated temperature probes ($\pm$ 0.05 $^{\circ}$C uncertainty) served as a test Rb vapor source providing known density values. For the densities used during the experiments (1 to $10\cdot10^{14}$ cm$^{-3}$), the systematic uncertainty is 0.3 $\%$ to 2 $\%$.\\
The two remaining tasks are online monitoring and controlling the experiment key parameters: the Rb vapor density and gradient. Here the important observable is the statistical uncertainty in density measurement and the systematic uncertainty between both diagnostics for equal densities. We determine it from data taken within a short amount of time (e.g. 5 min), knowing that the temperature in the source changes on much longer timescales ($\sim$ hours). This proves as well the ability of the analysis procedure to predict $n_{Rb}$ against variations in the other fitting parameter on short time scales (e.g. in $\Delta l$ due to vibrations).\\
 Figure \ref{fig:constdens} shows the Rb densities vs time at the AWAKE baseline density when the valves on top of the Rb reservoirs open and the 10 m pipe fills with Rb vapor. The density stabilized after $\approx \ 6$ min. After stabilization, one measures at the upstream end $n_{Rb,1}=(7.719\pm0.006)\cdot10^{14}$ cm$^{-3}$ ($\pm \ 0.08 \ \%$ standard deviation) and downstream $n_{Rb,2}=(7.715\pm0.007)\cdot10^{14}$ cm$^{-3}$ ($\pm \ 0.09 \ \%$ standard deviation), in both cases averaged over measurements taken over 5 min. The gradient over 10 m along the source is ($+ \ 0.05 \ \pm$ 0.12) $\%$.
 For the entire density range, the density values at constant temperature show statistical uncertainties between $\pm$ 0.05 $\%$ at high densities and $\pm$ 0.30 $\%$ at low densities.
 Combining this statistical uncertainties with the systematic error of $\pm$ 0.10 $\%$ to $\pm$ 0.35 $\%$ (checked in  \cite{MasterThesis}) found in measuring the same Rb vapor density at different locations with two independent diagnostics leads to a total uncertainty between both measurements of $\pm$ 0.11 $\%$ to $\pm$ 0.46 $\%$ (added in quadrature).\\
To study the effect of a density gradient on the SSM, we change the temperature of the downstream reservoir, but keep the density constant upstream. These temperature adjustments require live monitoring of the densities. Figure \ref{fig:gradscan} (a) shows an example of a such gradient scan, where the change in density was controlled and  adjusted based on the density values provided online by this diagnostic. Figure (b) shows the resulting change in the gradient. It was increased from 0 $\%$ to $\approx \ +20 \ \%$ over 10 m and then decreased to 6.70 $\%$ over 10 m (stable).

\section{Conclusion}
In conclusion, a method to measure Rb vapor densities in a fully automated way allowing for an online analysis is described. We use white light interferometry where two independent, fiber-based Mach-Zehnder interferometers measure the Rb vapor density at each end of the vapor source. Fourier-based signal conditioning and a fit algorithm retrieve the density values with an uncertainty between both measurements of $\pm$ 0.11 $\%$ to $\pm$ 0.46 $\%$. This precision fulfills the requirements to determine the density gradient over 10 m within 1 $\%$. This is the main diagnostic to monitor and control the plasma density and is crucial for an effective wakefield formation and electron acceleration.

\section*{Acknowledgments}
This work is sponsored by the Wolfgang Gentner Program of the German Federal Ministry of Education and Research (05E15CHA).

\nocite{Gschwendtner:1748353}
\nocite{metalpressure2}
\nocite{Fiberinterferometer}


\end{document}